\documentstyle[12pt,epsfig,ldd_art]{article} 
\input{epsf.tex}
\input{psfig.sty}
\newcommand{\beq}{\begin{equation}}
\newcommand{\eeq}{\end{equation}}
\newcommand{\bea}{\begin{eqnarray}}
\newcommand{\eea}{\end{eqnarray}}

\newcommand{\cond}{\langle\bar\chi\chi\rangle}

\newcommand{\oh}{\frac{1}{2}}

\newcommand{\non}{\nonumber}

\newcommand{\lapprox}{\raisebox{-0.5ex}{$\
\stackrel{\textstyle<}{\textstyle\sim}\ $}}

\newcommand{\One}{1\kern-4.5pt1}
\begin{document}

\addtolength{\baselineskip}{0.20\baselineskip}


\hfill SHEP/99/11

\hfill SWAT/99/213

\hfill February 1999

\begin{center}

\vspace{48pt}

{ {\bf The Three Dimensional Thirring Model for $N_f=4$ and $N_f=6$} }

\end{center}

\vspace{18pt}

\centerline{\sl L. Del Debbio~$^a$
and S.J. Hands~$^b$}
\centerline{\sl (the UKQCD collaboration)}

\vspace{15pt}

\centerline{$^a$Department of Physics and Astronomy, University of Southampton,}
\centerline{Highfield, Southampton SO17 1BJ, U.K.}
\vspace{15pt}

\centerline{$^b$Department of Physics, University of Wales Swansea,}
\centerline{Singleton Park, Swansea SA2 8PP, U.K.}

\vspace{48pt}

\begin{center}  

{\bf Abstract}

\end{center}

We present Monte Carlo simulation results for the three dimensional Thirring
model for numbers of fermion flavors $N_f=4$ and 6. For $N_f=4$ we 
find a second order chiral symmetry breaking transition at strong coupling, 
corresponding to an ultra-violet fixed point of the renormalisation group
defining a non-trivial continuum limit. The
critical exponents extracted from a fit to a model equation of state are
distinct from those found for $N_f=2$. For $N_f=6$, in contrast, we present
evidence for tunnelling between chirally symmetric and broken vacua at strong
coupling, implying that the phase transition is first order and no continuum
limit exists. The implications for the phase diagram of the model in the 
plane of coupling strength and $N_f$ are briefly discussed.



\vfill

\newpage

\section{Introduction}

The study of quantum field theories
in which the ground state shows a sensitivity to the number of light 
fermion flavors $N_f$
is intrinsically interesting. Examples include QCD-like theories with an 
intermediate number of flavors \cite{BZ}, $N=1$
supersymmetric QCD
\cite{Seiberg}, and the properties of QCD itself at high baryon number density
\cite{ARW}. Two model field theories which are thought to display
this phenomenon in three spacetime dimensions are QED 
\cite{ABKW} and the Thirring model 
[5-9].
$\mbox{QED}_3$, which is super-renormalisable, 
is believed to exist in a state of
spontaneously broken chiral symmetry for $N_f<N_{fc}$, where $N_{fc}$ is
some critical value. It has been suggested that the
infra-red behaviour is described by a 
{\sl conformal fixed point\/} \cite{MirYam}\cite{Gusy}, ie. that the critical
scaling exhibits an essential singularity as $N_f\nearrow N_{fc}$. 
Since infra-red properties are governed by
strongly coupled dynamics due to the model's asymptotic freedom, the
determination of $N_{fc}$ and description of the fixed point 
is an inherently non-perturbative problem. The Thirring model, 
by contrast, is non-renormalisable for $d>2$, but renormalisable in 
a large-$N_f$ expansion \cite{Gomes}\cite{largeN}, which predicts that the 
ground state has unbroken chiral symmetry. On the other hand, 
Schwinger-Dyson \cite{Itoh} and lattice \cite{DHM} studies
suggest that for $N_f$ less than some 
$N_{fc}$ at strong coupling chiral symmetry is spontaneously broken, 
the transition at the critical coupling $g_c^2$ defining an ultra-violet
renormalisation group (RG) fixed point. Once again, the issues of the
numerical value of $N_{fc}$ and the nature of the critical scaling must be
addressed by non-perturbative means.

It is natural to speculate whether $N_{fc}$ and
the critical behaviour of the two models
might be related. As we shall outline in the next section, 
the pattern of global symmetry breaking is the same in both cases, and hence 
one might naively expect the universality classes to coincide. Another 
suggestive argument is that the Schwinger-Dyson equation 
describing 
the IR behaviour of $\mbox{QED}_3$ 
is identical to that describing the UV of the Thirring
model at strong coupling \cite{Itoh}. 
One should be cautious, however, firstly 
because the results of Schwinger-Dyson
studies may be sensitive to the truncations employed, and secondly because
universality arguments may not apply in the presence of a massless particle, 
whose presence in the $\mbox{QED}_3$ spectrum is guaranteed by gauge
invariance, but which is only predicted in the Thirring model in the strong 
coupling limit. Nonetheless, it would be interesting if the UV fixed points
found for finite coupling in the Thirring model \cite{DHM} were related in 
any way to approximate IR fixed points of $\mbox{QED}_3$
invoked to account for non-Fermi liquid behaviour in the normal phase 
of high temperature superconductors \cite{Aitch}.

In previous lattice studies \cite{DHM} we have performed Monte Carlo
simulations of the Thirring model with $N_f=2$, 4 and 6 (the simulation
algorithm used requires $N_f$ to be even, as outlined in the next section).
For $N_f=2$ and 4 we found evidence for spontaneous chiral symmetry breaking at
strong coupling, and studies of the $N_f=2$ case
from a variety of lattice volumes and bare fermion
masses in the neighbourhood of the transition permitted a finite volume 
scaling analysis of the model's equation of state (EOS).
The result was that a continuous phase transition was found characterised
by a critical inverse coupling $1/g_c^2=1.92(2)$ and critical exponents 
$\delta=2.75(9)$, $\beta=0.57(2)$, $\nu=0.71(4)$, where certain assumptions 
such
as hyperscaling were used to extract the latter values. The implication is that
a continuum limit exists at the critical point, described by an interacting
quantum field theory. These results have recently been corroborated in an
independent study of the $\chi U\phi_3$ model \cite{BPFFJ}, 
a model of interacting scalars, fermions and gauge fields. In the strong
gauge coupling limit 
it can be shown that this model is equivalent to
our lattice  $N_f=2$ Thirring model, with the mapping \cite{FJ}
\begin{equation}
{1\over g^2}={{2r^2}\over{1-r^2}}\;\;\;\mbox{with}\;\;\;
r={{I_1(2\kappa)}\over{I_0(2\kappa)}},
\end{equation}
where $g^2$ is the Thirring coupling constant and $\kappa$ is the
hopping parameter of the scalar field in the $\chi U\phi_3$ model.
The strong coupling 
results of ref. \cite{BPFFJ}, based on fits to equations of state and
spectroscopy and a study of Lee-Yang zeros, 
and making different assumptions about the critical scaling, 
are $\kappa_c=0.983(12)\Rightarrow1/g_c^2=1.84(4)$,
$\delta=3.45(71)$, $\beta=0.51(11)$ and $\nu=0.75(10)$, which are compatible
with ours. We note in passing that the exponent $\tilde\nu$ of \cite{BPFFJ}
can be identified with the ratio $\nu/\Delta$, where 
$\Delta=\delta\beta$ is the gap exponent associated with the critical scaling
of the Lee-Yang edge singularity \cite{Kocic}.

In \cite{BPFFJ} the main thrust of the analysis was to search for possible new
RG fixed points in a coupling space of higher dimension, ie. away from the
strong gauge coupling limit, but keeping $N_f=2$.
In this paper we explore a different direction, namely 
the effect of increasing the number of fermion flavors, 
by extending  our earlier Monte Carlo 
simulations to encompass the cases
$N_f=4$ and $N_f=6$. Most of the new results we present will be from 
a $16^3$ lattice with bare fermion mass $m=0.01$ in 
lattice units, closer to the chiral limit than previous studies. 
We shall see, using an analysis
identical to that of \cite{DHM}, that for $N_f=4$ 
the data is well fitted by the
assumption of a critical equation of state at the chiral
transition, yielding exponent values distinct from those for $N_f=2$.
This is consistent with the scenario that both models define UV 
RG 
fixed points, described by distinct field theories, and that the critical
$N_{fc}>4$. For $N_f=6$, by contrast, no critical scaling is observed; instead
our data is consistent with there being a first order chiral symmetry breaking
phase transition, implying that in this case there is no continuum limit, 
and that therefore $N_{fc}<6$. In the next section we present the lattice
model in detail and review its pattern of symmetry breaking,
contrasting this with the continuum model. In section 3 
we present results from simulations of the $N_f=4$ model, including fits to an
RG-inspired equation of state, an attempt to construct a scaling function, and
details of the model's spectrum (both of the fundamental fermion and $f\bar f$ 
bound states) and susceptibilities, which enables an 
interesting comparison, both qualitative and quantitative, with $N_f=2$. 
The phenomenon of parity doubling in the
spin-1 sector, observed in \cite{DHM}, is also 
explained more fully here.
Section 4 concentrates on simulations of the $N_f=6$ model; here we show
evidence for metastability in the critical region, suggestive of a first order
transition. In section 5 we present a summary and conclusions.

\section{Lattice Formulation }

The lattice action we simulate employs the staggered fermion formulation,
with an auxiliary vector field $A_\mu$ defined on the lattice links 
\cite{DHM}:
\bea
        S &=& \oh \sum_{x\mu i} \bar\chi_i(x) \eta_\mu(x)
        \biggl[(1+iA_\mu(x)) \chi_i(x+\hat\mu)
        -(1-iA_\mu(x-\hat\mu))\chi_i(x-\hat\mu)\biggr] \non \\
          & & + m \sum_{xi} \bar\chi_i(x) \chi_i(x) +
        \frac{N}{4g^2} \sum_{x\mu} A_\mu^2(x). 
\label{eq:non-compact}
\eea
Here $\eta_\mu$ are the Kawamoto-Smit phases, and the index $i$ runs
over $N$ flavors of staggered fermion. The auxiliary field may be integrated
over to yield a form of the action with explicit four-fermion couplings:
\bea
        S &=& \oh \sum_{x\mu i}
 \bar\chi_i(x) \eta_\mu(x) \left[\chi_i(x+\hat\mu) -
        \chi_i(x-\hat\mu)\right] + 
          m \sum_{xi} \bar\chi_i(x) \chi_i(x) + \non \\
& &\frac{g^2}{4N} \sum_{x\mu ij} \biggl[2\bar\chi_i(x)\chi_i(x+\hat\mu)
        \bar\chi_j(x+\hat\mu)\chi_j(x)+\non\\
& &\bar\chi_i(x)\chi_i(x+\hat\mu)\bar\chi_j(x)\chi_j(x+\hat\mu)
+\bar\chi_i(x+\hat\mu)\chi_i(x)\bar\chi_j(x+\hat\mu)\chi_j(x)\biggr].
\label{eq:lattice-thirring}
\eea
Note that the last two four-fermi terms, which vanish for $N=1$ due to the
Grassmann nature of $\chi$, $\bar\chi$, were mistakenly omitted in eqn. (2.1)
of \cite{DHM}.

It is possible to rewrite the action (\ref{eq:non-compact}) in terms of
fields $q$, $\bar q$, which carry explicit spin and flavor indices
\cite{BB}\cite{DHM}. One then finds that the number of continuum four-component
fermions $N_f$ is related to $N$ via
\begin{equation}
N_f=2N.
\end{equation}
It is interesting, however, to compare the global symmetries of the lattice
action with those of the continuum Thirring model with $N_f$ flavors,
which are the same as $\mbox{QED}_3$ \cite{Gusy}. In the continuum model in the
chiral limit $m\to0$, there is a global symmetry generated by the $4\times4$
Dirac matrices $\One, \gamma_4,\gamma_5,\gamma_4\gamma_5$, which when combined 
with explicit flavor rotations means that the full global symmetry group
is $\mbox{U}(2N_f)$. The parity-invariant mass term $m\bar\psi\psi$ is not
invariant under rotations generated by either $\gamma_4$ or $\gamma_5$, but
leaves two independent $\mbox{U}(N_f)$ symmetries unbroken. The proposed
pattern of
chiral symmetry breaking in the continuum model is thus
\begin{equation}
\mbox{U}(2N_f)\to\mbox{U}(N_f)\otimes\mbox{U}(N_f).
\label{eq:contxsb}
\end{equation}
For the lattice action (\ref{eq:non-compact}) we identify a global symmetry in
the massless limit:
\bea
\bar\chi_o\mapsto\bar\chi_o U & & \chi_e\mapsto U^\dagger\chi_e \non \\
\bar\chi_e\mapsto\bar\chi_e V & & \chi_o\mapsto V^\dagger\chi_o,
\eea
where $\chi_{o/e}$ denotes the field defined on odd (ie.
$\varepsilon(x)=(-1)^{x_1+x_2+x_3}=-1$) 
and even sites respectively, and $U$, $V$ are
independent $\mbox{U}(N)$ matrices. With $m\not=0$, the symmetry only persists
for $U\equiv V$; hence the pattern of chiral symmetry breaking is
\begin{equation}
\mbox{U}(N)\otimes\mbox{U}(N)\to\mbox{U}(N).
\label{eq:latxsb}
\end{equation}

It is an open question whether there is a continuum limit of the lattice model
in which the pattern (\ref{eq:contxsb}) is approximately realised. This
could in principle be resolved in a simulation by careful analysis of, say,
the spectrum of approximate Goldstone modes. Another possibility, which 
must be given serious consideration due to the strongly-coupled nature
of any putative fixed point, is that 
it is the lattice pattern (\ref{eq:latxsb}) which characterises the continuum
limit, 
and that the form (\ref{eq:contxsb}) is not realised. In this scenario
the $N=1$ model would share the same 
symmetry breaking pattern (\ref{eq:latxsb})
as the lattice Gross-Neveu model with continuous chiral symmetry
considered in \cite{FJP}, with $N_f=2$ (ie.
$N=1$). The
smallest number of flavors for which this model has been simulated 
using a hybrid Monte Carlo algorithm is
$N_f=4$ ($N=2$)\cite{FJP}. An interesting possibility is that the lattice
versions of the Thirring and Gross-Neveu models lie in the same universality
class for $N_f=2$ \cite{BPFFJ}.
Finally, we note that the global symmetries of (\ref{eq:non-compact})
are identical to those of non-compact lattice $\mbox{QED}_3$.

In the work presented here we simulated the action (\ref{eq:non-compact})
using a standard hybrid Monte Carlo algorithm. The form of the action permits
an even-odd partitioning, so that there is no extra doubling of fermion
species. We will present new results here for the cases $N=2$ and $N=3$,
corresponding to $N_f=4$ and $N_f=6$ respectively. The measurements we
perform, and the nomenclature we use, 
are exactly the same as those described for the $N_f=2$ case in
\cite{DHM}, to which we refer the reader for technical details (although note
that the factors of $1/V$ appearing in the susceptibility definitions
(2.25-27) of \cite{DHM} are incorrect).
Most of the new results in this paper were obtained on a $16^3$ system with 
bare fermion mass $m=0.01$. It is worth recording the numerical effort involved,
since it is surprisingly large. To maintain a reasonable acceptance rate
in the hybrid Monte Carlo, we used timesteps typically between 0.01 and 0.015.
Our conjugate gradient routine was set to accept residual norms of
$10^{-6}$ per lattice site during guidance and $10^{-9}$ per site on the
Metropolis step: we found the number of iterations required varied from 600 in
the symmetric phase to 1500 in the broken phase during guidance, and from 800
to 1900 during the Metropolis step. A large amount of computational
effort is also
required in the $\beta=0$ limit of the $\chi U\phi_3$ model \cite{BPFFJ}.

\section{$N_f=4$}

This section is devoted to a complete analysis of the RG structure
of the theory for $N_f=4$. The approach adopted here has been
explained in detail in previous publications~\cite{DHM}.

\subsection{Fits to the Equation of State}
\label{subs:eos}

\begin{table}[ht]
\setlength{\tabcolsep}{1.5pc}
\caption{List of results for the chiral condensate for $N_f=4$}
\label{tab:new_data_N=4}
\begin{tabular*}{\textwidth}{@{}l@{\extracolsep{\fill}}rrrrr}
\hline
$L$ &  $m$  & $1/g^2$   & $\cond$ & $\Delta\cond$ \\
\hline
16 & 0.01 & 0.5  & 0.2199 & 0.0025 \\
16 & 0.01 & 0.6  & 0.1912 & 0.0025 \\
16 & 0.01 & 0.65 & 0.1717 & 0.0030 \\
16 & 0.01 & 0.67 & 0.1547 & 0.0030 \\
16 & 0.01 & 0.7  & 0.1403 & 0.0021 \\
16 & 0.01 & 0.75 & 0.1136 & 0.0025 \\
16 & 0.01 & 0.8  & 0.0903 & 0.0028 \\
16 & 0.01 & 0.9  & 0.0591 & 0.0018 \\
16 & 0.02 & 0.5  & 0.2473 & 0.0014 \\
16 & 0.02 & 0.6  & 0.2157 & 0.0015 \\
16 & 0.02 & 0.65 & 0.2011 & 0.0011 \\
16 & 0.02 & 0.67 & 0.1923 & 0.0012 \\
16 & 0.02 & 0.7  & 0.1774 & 0.0011 \\
16 & 0.02 & 0.8  & 0.1367 & 0.0015 \\
16 & 0.02 & 0.9  & 0.1056 & 0.0014 \\
16 & 0.03 & 0.65 & 0.2228 & 0.0008 \\
16 & 0.03 & 0.67 & 0.2135 & 0.0010 \\
16 & 0.03 & 0.7  & 0.2022 & 0.0009 \\
16 & 0.04 & 0.65 & 0.2358 & 0.0008 \\
16 & 0.04 & 0.67 & 0.2294 & 0.0008 \\
16 & 0.04 & 0.7  & 0.2208 & 0.0009 \\
\hline
\end{tabular*}
\end{table}

First results for $N_f=4$ were presented in~\cite{DHM}. Further
results from a $16^3$ lattice are analysed here by fitting to an
equation of state for fixed lattice size. Next, by combining the outcome of the
new runs with previously published results, a fit to an equation of
state including finite size effects is also presented.
For the sake of completeness, the main results leading to the equation
of state are summarised.

\begin{figure}[htb]
\psdraft
\centerline{
\setlength\epsfxsize{300pt}
\epsfbox{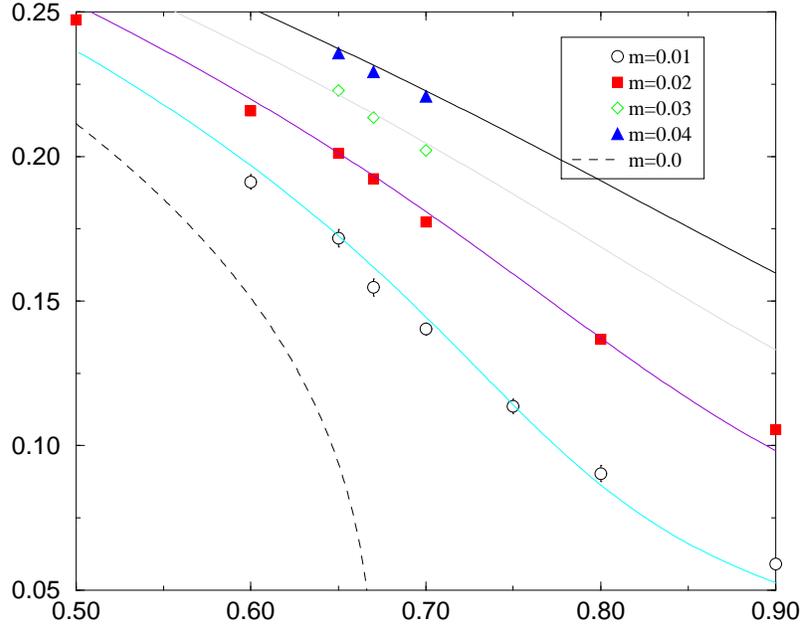
}}
\psfull
\caption{Chiral condensate vs. $1/g^2$ on $16^3$ lattice.
The solid lines are the fits from (\ref{eq:eos2}), the dashed line
the extrapolation to the chiral limit.
\label{fig:chi16}}
\end{figure}

For fixed lattice size, the solution of the RG equation in a neighbourhood of
a fixed point yields a generic relation between the order parameter
and the external symmetry breaking
field, which is called an equation of state:
\begin{equation}
        m\left(\langle \bar\chi \chi \rangle,t,1\right) \sim 
	\langle \bar\chi \chi \rangle^\delta 
	{\cal F}(t \langle \bar\chi \chi \rangle^{-1/\beta}),
\label{eq:scaling1}
\end{equation}
where for the Thirring model the order parameter is the chiral condensate
$\langle \bar\chi \chi \rangle$, the symmetry breaking field the 
bare fermion mass $m$, and the reduced coupling $t$ parametrising the distance
from criticality identified with 
\begin{equation}
        t = 1/g^2 - 1/g_c^2.
\label{eq:redt}
\end{equation}
$\cal F$ is a universal scaling function. By setting $m=0$ in
Eq.~(\ref{eq:scaling1}), the critical behaviour of the order parameter
when the external field is switched off is recovered:
\begin{equation}
        t \langle \bar\chi \chi \rangle^{-1/\beta} \sim \mbox{const}
\end{equation}
while, for $t=0$, ${\cal F}(0)$ is a constant and hence:
\begin{equation}
        m \sim \langle \bar\chi \chi \rangle^{\delta}
\end{equation}
showing clearly that $\beta$ and $\delta$ are the usual critical
exponents introduced in the context of phase transitions.  If the
critical exponents are related to the existence of a UV fixed point,
as we are assuming in this section, they must obey the hyper-scaling
relations:
\begin{eqnarray}
        \beta &=& \oh \nu (d-2+\eta) \label{eq:beta_hs} \\
        \delta &=& \frac{d+2-\eta}{d-2+\eta}
\label{eq:beta}
\end{eqnarray}
where $\eta$ is related to the anomalous dimension of $\bar\chi\chi$
and $\nu$ is the critical
exponent which characterises the divergence of the correlation length
as $t\rightarrow 0$.

A Taylor expansion for small $t$ reduces Eq.~(\ref{eq:scaling1}) to
an expression which can be used to fit the lattice data:
\begin{equation}
        m = B \cond^\delta + A t \cond^{\delta-1/\beta} +
        {\cal O}\left((t\cond^{-1/\beta})^2\right).
\label{eq:eos1}
\end{equation}
The new set of data generated on the $16^3$ lattice are summarised in
Tab.~\ref{tab:new_data_N=4}.

Since Eq.~(\ref{eq:eos1}) is obtained from a Taylor expansion around the
critical coupling, it is only expected to fit the data in a close
neighbourhood around the latter. The number of data points included in
the fit is chosen in order to minimize the $\chi^2$/d.o.f.

\begin{table}[ht]
\setlength{\tabcolsep}{1.5pc}
\caption{Results for $N_f=4$ from fits on the $12^3$ and $16^3$
lattices.}
\label{tab:fit_results}
\begin{tabular*}{\textwidth}{@{}l@{\extracolsep{\fill}}rrrr}
\hline
        & Parameter      &   Fit I     &   Fit II     \\
\hline
$16^3$  & $1/g^2_c$      &             &  0.67(9)     \\
        & $\delta$       &             &  3.64(18)    \\
        & $\beta$        &             &  ---         \\
        & $A$            &             &  0.837(2)    \\
        & $B$            &             &  8.61(1.9)   \\
        & $\chi^2$/d.o.f &             &  2.3         \\
        &                &             &              \\
$12^3$  & $1/g^2_c$      &  0.63(1)    &  0.66(1)     \\
        & $\delta$       &  3.67(28)   &  3.43(19)    \\
        & $\beta$        &  0.38(4)    &  ---     \\
        & $A$            &  0.78(5)    &  0.73(2)     \\
        & $B$            &  7.9(2.8)   &  6.4(1.5)    \\
        & $\chi^2$/d.o.f &  3.1        &  2.0         \\
\hline
\end{tabular*}
\end{table}

The results of the fit together with previously published results are
reported in Tab.~\ref{tab:fit_results}.  Fit I is a fit in which 
both $\delta$ and $\beta$ are kept as free parameters; fit II
imposes the constraint $\delta-1/\beta=1$, originally
inspired by Schwinger-Dyson solutions of the gauged Nambu -- Jona-Lasinio
model \cite{DHKK}, and consistent with the degeneracy of
scalar and pseudoscalar bound states in the chirally symmetric phase
\cite{DHM}. The agreement with the
previous results on the $12^3$ lattice indicates both that the lattice
sizes considered here are sufficiently close to the infinite volume
limit, and that this additional constraint is approximately 
obeyed by the data. 
It is therefore possible to try to include finite size effects
in the equation of state following the prescription presented
in~\cite{DHM}.

The inverse size of the lattice, in units of the lattice spacing, can
be included in the RGE as a relevant coupling with eigenvalue 1, with
a fixed point at $1/L=0$~\cite{zinn90}. The equation of state obtained
in this framework is~\cite{DHM}:
\begin{equation}
        m \cond^{-\delta} \sim  {\cal F}(t \cond^{-1/\beta},
        L^{-1/\nu} \cond^{-1/\beta})
\label{eq:scaling2}
\end{equation}
where ${\cal F}$ is now a universal scaling function of two rescaled
variables. The data are then fitted to the equation obtained by
Taylor expansion of Eq.~(\ref{eq:scaling2}):
\begin{equation}
        m = B \cond^\delta + A
(t + C L^{-1/\nu}) \cond^{\delta-1/\beta} + \mbox{higher order terms},
\label{eq:eos2}
\end{equation}

\begin{table}[ht]
\caption{Results for $N_f=4$ from fit including finite size scaling.}
\label{tab:fss_fit}
\begin{minipage}{\linewidth}
\renewcommand{\thefootnote}{\thempfootnote}
\begin{tabular*}{\textwidth}{@{}l@{\extracolsep{\fill}}rrr}
\hline
        & Parameter      &   Fit III   \\
\hline
$N_f=4$ & $1/g^2_c$      &  0.69(1)    \\
        & $\delta$       &  3.76(14)    \\
        & $\beta$~\footnote{evaluated from
$\delta-1/\beta=1$ constraint}
                         &  0.36(2)    \\
        & $\eta$~\footnote{evaluated from hyperscaling relation}
                         &  0.26(4)    \\
        & $\nu$~\footnotemark[\value{mpfootnote}]
                         &  0.57(1)    \\
        & $A$            &  0.83(2)   \\
        & $B$            &  10(2)    \\
        & $C$            &  2.0(5)     \\
        & $\chi^2$/d.o.f &  2.0       \\
\hline
\end{tabular*}
\end{minipage}
\end{table}

The results of the fit are shown in Tab.~\ref{tab:fss_fit}.  They
are consistent with those coming from the fixed size analysis,
confirming the existence of a fixed point, with non-gaussian critical
exponents. A non-trivial check comes from the value of $\beta$, which
is determined using the constraint $\delta - 1/\beta=1$, but must also
obey the hyperscaling relation. Plugging the fitted values for
$\delta$ and $\nu$ in Eq.~(\ref{eq:beta_hs}) yields $\beta=0.36$, in
agreement with the determination mentioned above.

The data for the fermion condensate as a function of the inverse
coupling are reported in Fig.~\ref{fig:chi16} for different values of
the bare mass. The dashed line represents the critical curve for
$m=0.0$ which is obtained from Eq.~(\ref{eq:eos2}) with $L=16$ and the
values obtained from the fit for the critical exponents. The solid lines
through the data points are also obtained from Eq.~(\ref{eq:eos2}). It
can be seen from the picture that the equation of state provides a
satisfactory description of the lattice data.

\subsection{Scaling function}
\begin{figure}[htb]
\centerline{
\setlength\epsfxsize{300pt}
\setlength\epsfysize{250pt}
\epsfbox{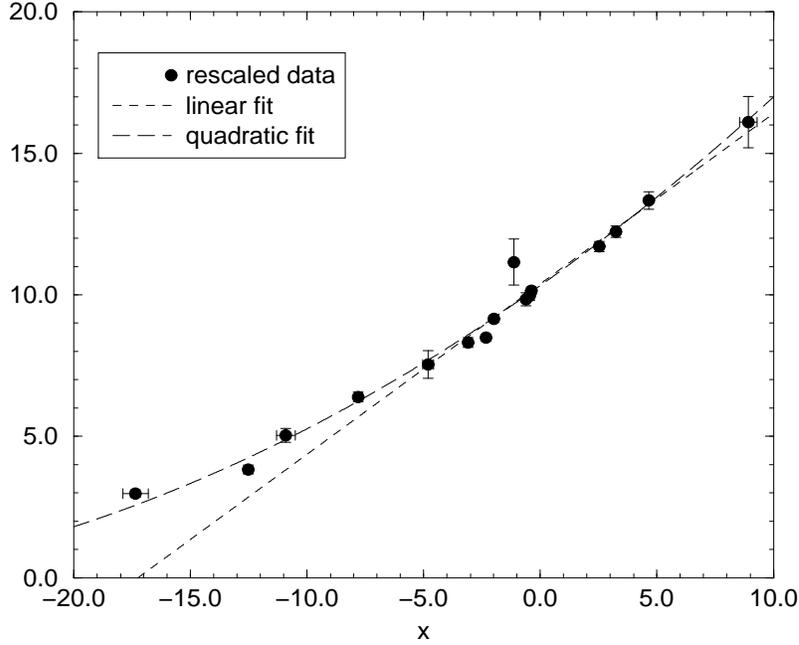}
}
\caption{The universal scaling function for $N_f=4$ reconstructed from
rescaled data, using the critical exponents determined from fit III
above.
\label{fig:scalf}}
\end{figure}
\begin{figure}[htb]
\centerline{
\setlength\epsfxsize{300pt}
\setlength\epsfysize{250pt}
\epsfbox{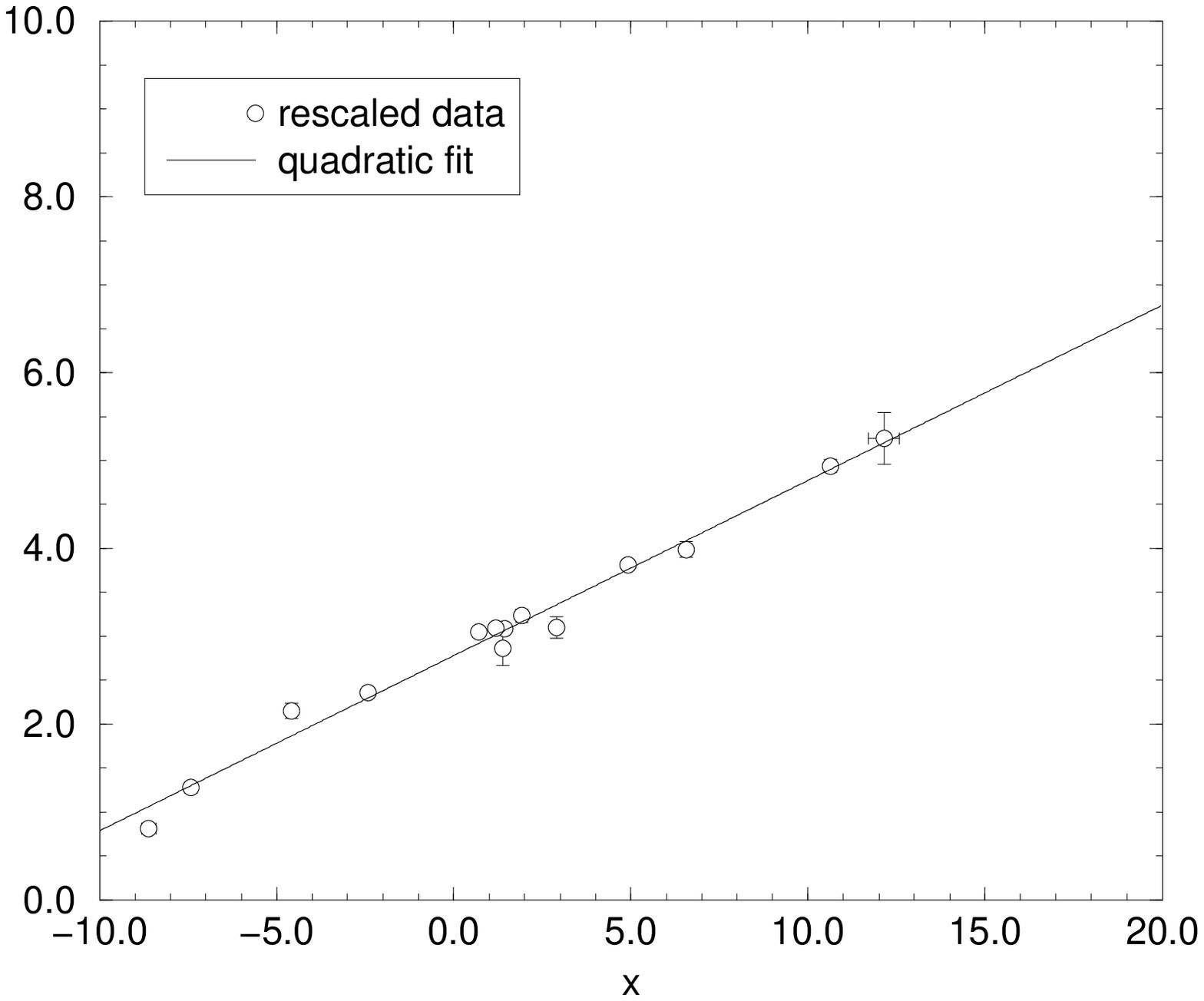}
}
\caption{The universal scaling function for $N_f=2$ reconstructed from
rescaled data, using the critical exponents determined from fit III
presented in \cite{DHM}.
\label{fig:scalf1}}
\end{figure}

The general form of the equation of state presented in
Eq.~(\ref{eq:scaling1}) suggests a further check of the values obtained
for the critical exponents~\cite{jersak97}. A plot of the rescaled
variables $m \cond^{-\delta}$ vs. $t \cond^{-1/\beta}$ should show that
the data from runs at different values of the coupling and the bare mass
lie on a single curve, describing the universal scaling function
${\cal F}$.

The curve shown in Fig.~\ref{fig:scalf} is obtained using the critical
exponents determined from the fit III to rescale the data points from
the $16^3$ lattice .  The points do indeed lie on a single curve
(within 1-2 standard deviations). Defining $x=t \cond^{-1/\beta}$, we
performed two different fits for the scaling function.
\begin{itemize}
\item a quadratic fit, which yields:
\end{itemize}
\begin{equation}
{\cal F}(x) = 10.329 + 0.586 x + 0.008 x^2
\end{equation}
The constant term agrees very well with the coefficient $B$ from fit
III, while the coefficient of the linear term is within 25\% of $A$. The
fact that the coefficient of the quadratic term turns out to be so
small confirms that the data are approximately described by a linear
function, providing yet more evidence in favour of our hypothesis
$\delta - 1/\beta=1$.
\begin{itemize}
\item on the range shown in Fig.~\ref{fig:scalf}, we also tried to fit
to the form:
\end{itemize}
\begin{equation}
{\cal F}(x) = p + q (20+x)^r
\end{equation}
in order to check for a possible non-linear behaviour. The curvature
which is visible in the data is reflected in the result of the fit:
$p=1.33$, $q=0.2$ and $r=1.28$. However, if we expand the result
around $x=0$, where we have fitted to the Taylor expansion of the EOS,
we obtain:
\begin{equation}
{\cal F}(x) = 10.472 + 0.584 x + 0.004 x^2
\end{equation}
which shows a very good agreement with the previous fit and the same
small coefficient for the quadratic term.

Those results confirm that our determination of the critical exponents
does allow one to rescale the data points on a single universal curve
and that the constraint $\delta-1/\beta=1$ provides a satisfactory
description of the data close to the critical point.

In order to check our previous determination of the critical exponents for the
$N_f=2$ case, the results of the same analysis on the old set of data
\cite{DHM} is presented in Fig.~\ref{fig:scalf1}. The results are qualitatively
similar. The critical exponents determined from the fit to the EOS
define the rescaled variables so that all the points are on a
universal curve. The outcome of the fit to the universal function
yields once again a very small quadratic coefficient.

\subsection{Susceptibilities}

\begin{table}[ht]
\caption{Susceptibilities for the $N_f=2$ model from a $16^3$ lattice
with $m=0.01$}
\label{tab:suscN=1}
\begin{minipage}{\linewidth}
\renewcommand{\thefootnote}{\thempfootnote}
\begin{tabular*}{\textwidth}{@{}l@{\extracolsep{\fill}}ll}
\hline
$1/g^2$ & $\chi_l$      & $\chi_t$ \\
\hline
$1.6 $  & 5.35(70) & 20.18(50)   \\
$1.8 $  & 5.44(74) & 14.25(40)   \\
$1.88$  & 5.23(31) & 12.50(21)   \\
$1.9 $  & 4.90(43) & 12.49(24)   \\
$1.92$  & 5.32(42) & 12.40(24)   \\
$2.0 $  & 5.37(48) & 10.25(50)   \\
$2.1 $  & 5.88(29) & 8.61(20)    \\
$2.2 $  & 5.88(33) & 7.18(12)    \\
$2.4 $  & 4.66(21) & 5.29(10)    \\
$3.0 $  & 3.05(13) & 3.10(3)     \\
\hline
\end{tabular*}
\end{minipage}
\end{table}

\begin{table}[ht]
\caption{Susceptibilities for the $N_f=4$ model from a $16^3$ lattice
with $m=0.01$}
\label{tab:suscN=2}
\begin{minipage}{\linewidth}
\renewcommand{\thefootnote}{\thempfootnote}
\begin{tabular*}{\textwidth}{@{}l@{\extracolsep{\fill}}ll}
\hline
$1/g^2$ & $\chi_l$      & $\chi_t$ \\
\hline
$0.5 $  & 1.93(75) & 21.80(29)  \\
$0.6 $  & 3.76(38) & 18.96(24)  \\
$0.65$  & 4.39(56) & 17.17(30)   \\
$0.67$  & 4.14(49) & 15.47(30)  \\
$0.7 $  & 3.35(50) & 14.03(21)  \\
$0.75$  & 5.53(32) & 11.36(25)  \\
$0.8 $  & 5.47(34) & 9.03(28)   \\
$0.9 $  & 5.18(24) & 5.91(18)   \\
\hline
\end{tabular*}
\end{minipage}
\end{table}

In this subsection we report on measurements of 
integrated two-point functions in scalar and pseudoscalar channels, 
which we denote by analogy with ferromagnetic systems as respectively
the longitudinal susceptibility $\chi_l$ and transverse susceptibility
$\chi_t$. It turns out that the susceptibilities
yield the most convincing evidence that the
scaling properties at the fixed points of the $N_f=2$ and $N_f=4$
models are distinct, and so we also present and plot results
from equivalent measurements for $N_f=2$.
In terms of the fermion kinetic operator $M$, $\chi_l$ is defined
by
\begin{eqnarray}
\chi_l & = & \left[\langle(\mbox{tr}M^{-1})^2\rangle-\langle\mbox{tr}M^{-1}
\rangle^2\right] -\sum_y\langle M^{-1}_{xy}M^{-1}_{yx}\rangle \nonumber \\
&\equiv& \chi_{ls} + \chi_{lns},
\end{eqnarray}
where we distinguish between a flavor singlet contribution given 
by diagrams formed from disconnected fermion lines, which
must be measured using a stochastic estimator, and a non-singlet 
contribution formed from diagrams used in standard meson spectroscopy,
with the source $x$ used for calculating the inverse of $M$ chosen 
at random for each measurement.
The transverse susceptibility has vanishing singlet part, and is given 
by
\begin{equation}
\chi_t=\sum_x\langle\varepsilon(0)M^{-1}_{0x}\varepsilon(x)M^{-1}_{x0}\rangle
\equiv{1\over m}\langle\bar\chi\chi\rangle,
\end{equation}
where the second equality results from the axial Ward identity.
In practice the Ward identity yields much the less noisy signal, and
is the one tabulated, though we 
have checked that the two relations give consistent results.

\begin{figure}[htb]
\psdraft
\centerline{
\setlength\epsfxsize{300pt}
\epsfbox{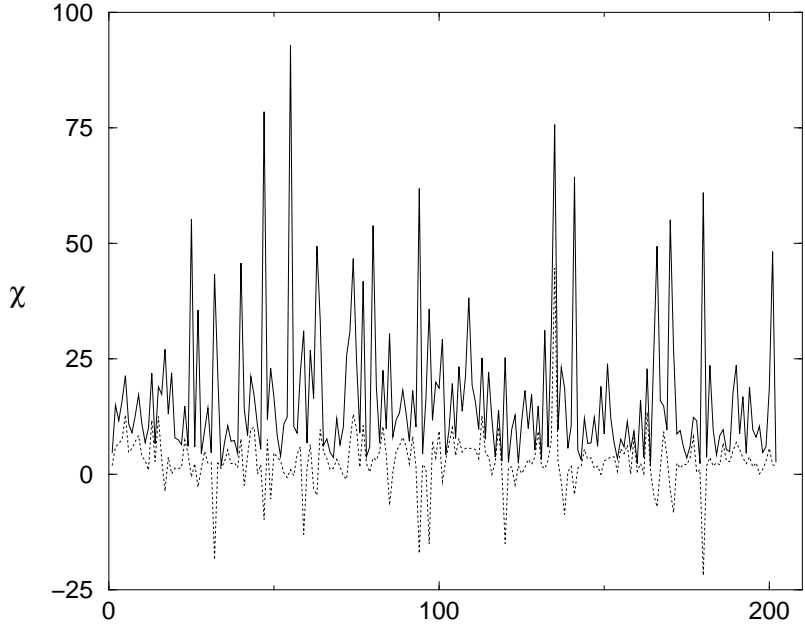
}}
\psfull
\caption{Estimates for $\chi_t$ (solid line) and $\chi_{lns}$ (dotted line)
for 200 measurements of the $N_f=4$ model with $m=0.01$, $1/g^2=0.67$.
\label{fig:history}}
\end{figure}

In Tables \ref{tab:suscN=1} and \ref{tab:suscN=2} we give our results
for susceptibilities obtained from $16^3$ lattices with $m=0.01$ for both
$N_f=2$ and $N_f=4$ (the $N_f=2$ results were plotted in \cite{DHM}).
It is interesting to note that the dominant source of statistical error in 
$\chi_l$,
particularly in the broken phase, comes from the non-singlet contribution
$\chi_{lns}$. The reason for this can be gleaned from inspecting a 
time history of the connected fermion line contribution to 
both $\chi_{lns}$ and $\chi_t$, as shown in Fig.~\ref{fig:history}.
Although the bulk of the $\chi_t$ values obtained are $\leq25$, there 
are a few measurements which yield significantly larger upward excursions
with correlated negative excursions for the estimate of $\chi_{lns}$.

A plausible explanation for this observation can be found by representing 
the fermion propagator in terms of the eigenmodes of 
$D{\!\!\!\!/}\,_{xy}\equiv M_{xy}-m\delta_{xy}$:
\begin{equation}
M^{-1}_{xy}=\sum_n{{\phi_n(x)\phi^*_n(y)}\over{i\lambda_n+m}},
\end{equation}
where the sum runs over the eigenmodes $\phi_n$ satisfying
$D{\!\!\!\!/}\,\phi_n(x)=i\lambda_n\phi_n(x)$. A straightforward calculation
then yields for the connected fermion line contribution evaluated with 
source $x$ the following:
\begin{equation}
\chi_{t/lns}(x)=\sum_{n}\phi^*_n(x)\phi_n(x)
{{\lambda_n^2\pm m^2}\over{(\lambda_n^2+m^2)^2}}.
\end{equation}
Hence we may attribute the large spikes in Fig. \ref{fig:history}
to configurations where there
is a particularly small eigenvalue $\vert\lambda\vert<m$ and the associated
eigenmode has a large value at the site of the source; the largest
upward excusions will thus be $\lapprox O(m^{-2})$.
The spikes make the signal noisy; for the data shown in 
Fig.~\ref{fig:history} the estimate for $\chi_t$ is $16.3\pm1.0$, whereas
the estimate from the Ward identity, $15.5\pm0.3$, is consistent but
with a much smaller error.

\begin{figure}[ht]
\psdraft
\centerline{
\setlength\epsfxsize{300pt}
\epsfbox{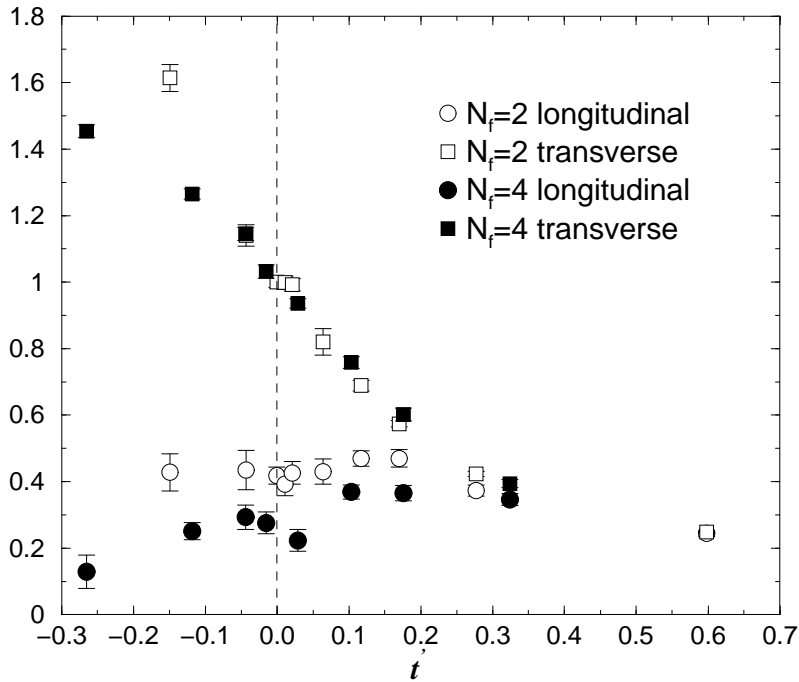
}}
\psfull
\caption{Reduced susceptibilities for $N_f=2,4$ from a $16^3$ lattice 
with $m=0.01$
\label{fig:reduced}}
\end{figure}

In order to make a meaningful comparison between susceptibilites
from $N_f=2,4$ we plot the data from both models as a function of
a reduced coupling $t^\prime$ distinct from that of
subsection \ref{subs:eos}, defined by
\begin{equation}
t^\prime=g_c^{\prime2}\left({1\over g^2}-{1\over g_c^{\prime2}}\right),
\end{equation}
with $g_c^{\prime2}$ chosen to be the value of $g^2$ for which 
the coefficient $(t+CL^{-1/\nu})$ vanishes in eqn. (\ref{eq:eos2}), 
using the fitted parameters of Table~\ref{tab:fss_fit} and Table 8 of
ref. \cite{DHM}. The data, which is also normalised such that
$\chi_t(t^\prime=0)=1$, is plotted in Fig.~\ref{fig:reduced}.
What we find is a marked difference between the shapes of the curves 
for the two models, even once the admittedly large errors are taken into
account; the $N_f=2$  data show $\chi_l$ and $\chi_t$
almost degenerate deep in the symmetric phase, and then $\chi_t$ increasing
with positive curvature into the broken phase while $\chi_l$ remains
roughly constant. The $N_f=4$ data, by contrast, suggest that the rise
of $\chi_t$ in the broken phase is less steep, and that at the same
time $\chi_l$ decreases.

\begin{table}[ht]
\caption{Susceptibilities and $R_\pi$ for a range of masses 
from a $16^3$ lattice}
\label{tab:Rpi}
\begin{minipage}{\linewidth}
\renewcommand{\thefootnote}{\thempfootnote}
\begin{tabular*}{\textwidth}{@{}l@{\extracolsep{\fill}}lllll}
\hline
$N_f$ & $1/g^2$ & $m$ & $\chi_l$ & $\chi_t$ & $R_\pi$ \\
\hline\hline
2 & 1.88 & 0.01 & 5.23(31)  & 12.50(21) & 0.418(26)  \\
  &      & 0.02 & 2.91(22)  &  8.28(8)  & 0.351(26)  \\
  &      & 0.03 & 2.33(15)  &  6.41(5)  & 0.363(23)  \\
  &      & 0.04 & 1.77(10)  &  5.24(3)  & 0.337(19)  \\
\hline
4 & 0.65 & 0.01 & 4.39(56)  &  17.17(30) & 0.256(33)  \\
  &      & 0.02 & 1.63(20)  &  10.06(9)  & 0.162(20)  \\
  &      & 0.03 & 1.21(13)  &  7.43(5)   & 0.162(17)  \\
\medskip
  &      & 0.04 & 0.89(9)   &  5.90(3)   & 0.151(15)  \\
  & 0.67 & 0.01 & 4.14(49)  &  15.47(30) & 0.267(32)  \\
  &      & 0.02 & 2.04(19)  &  9.62(10)  & 0.212(20)  \\
  &      & 0.03 & 1.56(13)  &  7.12(5)   & 0.219(18)  \\
\medskip
  &      & 0.04 & 0.86(12)  &  5.74(4)   & 0.149(20)  \\
  & 0.70 & 0.01 & 3.35(50)  &  14.03(21) & 0.239(35)  \\
  &      & 0.02 & 2.16(16)  &  8.87(11)  & 0.243(19)  \\
  &      & 0.03 & 1.49(15)  &  6.74(5)   & 0.221(22)  \\
  &      & 0.04 & 1.23(10)  &  5.52(5)   & 0.224(18)  \\
\hline
\end{tabular*}
\end{minipage}
\end{table}

\begin{figure}[htb]
\psdraft
\centerline{
\setlength\epsfxsize{300pt}
\epsfbox{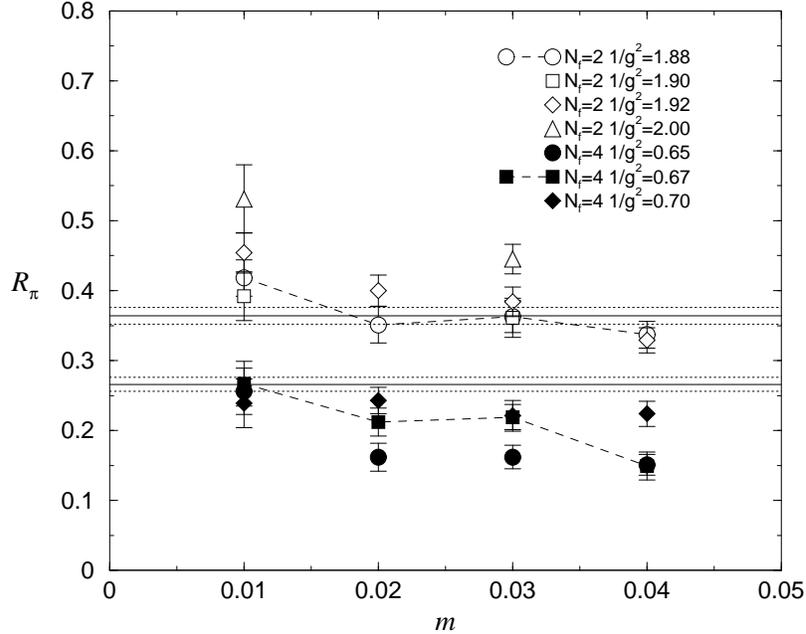
}}
\psfull
\caption{The ratio $R_\pi$ vs. $m$ for $N_f=2,4$ from a $16^3$ lattice.
\label{fig:Rpi}}
\end{figure}
The results of Fig.~\ref{fig:reduced}, being obtained away from the
chiral limit, can serve at best as a qualitative indicator of the differences
between the two models. A more quantitative guide comes from considering the
ratio $R_\pi\equiv\chi_l/\chi_t$ as a function of bare mass $m$
\cite{Rpi}\cite{DHM}. In general $R_\pi$ varies with $m$, but exactly
at the critical coupling, the equation of state (\ref{eq:eos1}) predicts
that $R_\pi=\delta^{-1}$ independent of $m$. In Table \ref{tab:Rpi}
we present results for susceptibilities and $R_\pi$ for $m=0.01,\ldots,0.04$
for a range of couplings in the critical region (including some
new results for $N_f=2$ at $1/g^2=1.88$); they are
plotted together with results
from Table 9 of ref. \cite{DHM} in Fig.~\ref{fig:Rpi}. Note
that over the mass range explored $\chi$ varies by a 
factor of $O(3)$. Also shown in the figure
are the values of $\delta^{-1}$ obtained from the equation of state
fits for $N_f=2,4$.
Within the large errors, we see that the results for $N_f=2$ with
$1/g^2=1.88$ (corresponding to $t^\prime\simeq0$) are roughly independent of 
$m$ and fall within the error band of $\delta$;
for $N_f=4$ both
independence of $m$ and agreement with $\delta$ is less convincing
for the $1/g^2=0.67$ data (ie. $t^\prime\simeq-0.015$), though still
plausible (it is also possible that the equation of state fits
have under-estimated $\delta$ in this case).
Whilst considerably more accuracy would be required
before this technique became a competitive means of estimating $1/g_c^2$,
in the critical region the values of $R_\pi$ for $N_f=2,4$
are clearly distinct, supporting our claim that the models' continuum limits
belong to different universality classes. One feature of Fig.~\ref{fig:Rpi}
that remains to be explained is the systematically larger values af $R_\pi$
for $m=0.01$: this could be either a finite volume effect, or an artifact due 
to insufficient sampling of the small eigenvalue configurations noted in
Fig.~\ref{fig:history}. Note that if such configurations are undersampled, 
the ratio $R_\pi$ will be {\sl over\/}-estimated.

\subsection{Spectrum}
The mass spectrum of the theory is studied by fitting the
time-dependence of the two-point functions with a single exponential
decay. The fermion propagator has been fitted to:
\begin{equation}
C_f(t)=A\left(e^{-\mu_Rt}-(-1)^t e^{-\mu_R(L-t)}\right),
\end{equation}
where $\mu_R$ is the physical fermion mass and
the minus sign between the forward and backward terms is due to our
choice of antiperiodic boundary conditions in the timelike direction.
Both the scalar and pion channels were fitted by the form
\begin{equation}
C_l(t)=A\left(e^{-Mt}+e^{-M(L-t)}\right).
\end{equation}

Although the lattice data has a larger statistical noise, the masses
in the three channels studied here show similar behaviours to those
obtained for $N_f=2$ in previous publications~\cite{DHM}.

The results are summarised in Tabs.~\ref{tab:spec_ferm},
\ref{tab:spec_pi} and \ref{tab:spec_scal}. The number of
configurations available for each value of $1/g^2$ and $m$ is reported
in square brackets.

\begin{table}[ht]
\caption{Results for the fermion spectrum.}
\label{tab:spec_ferm}
\begin{minipage}{\linewidth}
\renewcommand{\thefootnote}{\thempfootnote}
\begin{tabular*}{\textwidth}{@{}l@{\extracolsep{\fill}}rrrr}
\hline
$1/g^2$ & $m=0.01$      & $m=0.02$       & $m=0.03$      & $m=0.04$ \\
\hline
$0.5 $  & 0.65(9) [319] & 0.97(20) [107] &               &          \\
$0.6 $  & 0.46(5) [338] & 0.43(10) [105] &               &  \\
$0.65$  & 0.42(4) [210] & 0.50(5)  [200] & 0.48(3) [234] & 0.57(3) [208] \\
$0.67$  & 0.25(7) [206] &                & 0.45(4) [228] & 0.61(5) [219] \\
$0.7 $  & 0.32(3) [229] & 0.37(3)  [240] & 0.36(2) [250] & 0.43(2) [208] \\
$0.75$  & 0.16(3) [231] &                &               &          \\
$0.8 $  & 0.16(2) [205] & 0.23(2)  [102] &               &            \\
$0.9 $  & 0.06(1) [209] & 0.14(1)  [99]  &               &          \\
\hline
\end{tabular*}
\end{minipage}
\end{table}

\begin{table}[ht]
\caption{Results for the pion spectrum.}
\label{tab:spec_pi}
\begin{minipage}{\linewidth}
\renewcommand{\thefootnote}{\thempfootnote}
\begin{tabular*}{\textwidth}{@{}l@{\extracolsep{\fill}}rrrr}
\hline
$1/g^2$ & $m=0.01$       & $m=0.02$       & $m=0.03$        & $m=0.04$ \\
\hline
$0.5 $  & 0.164(5) [313] & 0.237(6) [107] &                 &          \\
$0.6 $  & 0.170(5) [333] & 0.234(8) [105] &                 &          \\
$0.65$  & 0.191(4) [209] & 0.258(5) [200] & 0.308(6)  [234] & 0.350(3) [208] \\
$0.67$  & 0.194(4) [206] & 0.264(4) [200] & 0.310(10) [228] & 0.353(4) [219] \\
$0.7 $  & 0.198(3) [221] & 0.261(5) [230] & 0.320(30) [250] &          \\
$0.75$  & 0.191(3) [228] &                &                 &          \\
$0.8 $  & 0.205(2) [201] & 0.272(7) [102] &                 &          \\
$0.9 $  & 0.223(1) [203] & 0.263(7) [99]  &                 &          \\
\hline
\end{tabular*}
\end{minipage}
\end{table}

\begin{table}[htb]
\caption{Results for the scalar spectrum.}
\label{tab:spec_scal}
\begin{minipage}{\linewidth}
\renewcommand{\thefootnote}{\thempfootnote}
\begin{tabular*}{\textwidth}{@{}l@{\extracolsep{\fill}}rrrr}
\hline
$1/g^2$ & $m=0.01$       & $m=0.02$       & $m=0.03$      & $m=0.04$ \\
\hline
$0.5 $  & 0.29(11) [305] & 0.34(11) [107] &               &          \\
$0.6 $  & 0.24(3)  [306] &                &               &          \\
$0.65$  & 0.28(7)  [209] & 0.55(13) [200] & 0.77(8) [225] & 0.86(5) [208] \\
$0.67$  & 0.29(8)  [202] & 0.54(11) [200] & 0.77(8) [192] & 0.92(7) [219] \\
$0.7 $  & 0.36(15) [213] & 0.48(10) [215] &               & 0.80(10)[208] \\
$0.75$  &                &                &               &          \\
$0.8 $  & 0.25(1)  [201] & 0.43(6)  [102] &               &          \\
$0.9 $  & 0.24(1)  [203] & 0.32(3)  [99]  &               &          \\
\hline
\end{tabular*}
\end{minipage}
\end{table}

\begin{table}[htb]
\caption{Results for the vector spectrum for $m=0.01$.}
\label{tab:spec_vect}
\begin{minipage}{\linewidth}
\renewcommand{\thefootnote}{\thempfootnote}
\begin{tabular*}{\textwidth}{@{}l@{\extracolsep{\fill}}ll}
\hline
$1/g^2$ & local current  & conserved current   \\
\hline
$0.9 $  & 0.34(8)  [203] & 0.31(4)  [203]        \\
$0.8 $  & 0.97(46)  [201] & 0.60(14)  [201]        \\
$0.75 $  & 0.31(17)  [210] & 0.50(11)  [204]        \\
$\leq0.7$ & no fit found & no fit found \\
\hline
\end{tabular*}
\end{minipage}
\end{table}

Although the signal for the fermion propagator is more noisy than in the
$N_f=2$ case, it is still possible to identify a clear increase in the
fermion mass when going from the symmetric to the broken phase of the
theory. The scalar propagator exhibits a very poor signal and it is often
difficult to get a stable result from the fit; even with this 
admittedly crude accuracy it is possible to see the transition between 
the chirally symmetric phase where the pion and scalar are approximately
degenerate, to the broken phase where the scalar is much heavier.

\begin{figure}[htb]
\psdraft
\centerline{
\setlength\epsfxsize{300pt}
\setlength\epsfysize{250pt}
\epsfbox{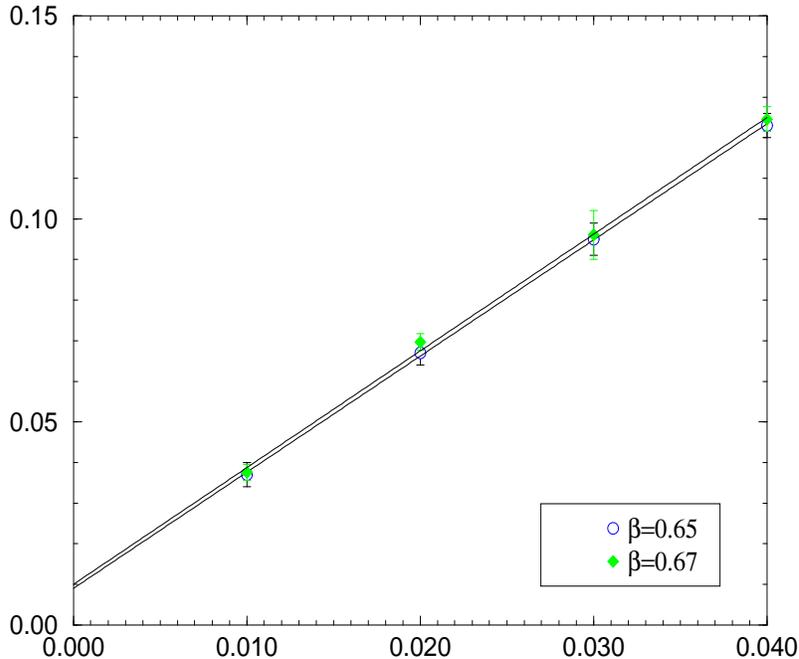}
}
\psfull
\caption{$M_\pi^2$ vs. $m$, for two different values of
$\beta$ in the broken phase.
\label{fig:mpivm}}
\end{figure}

\begin{figure}[htb]
\psdraft
\centerline{
\setlength\epsfxsize{300pt}
\setlength\epsfysize{250pt}
\epsfbox{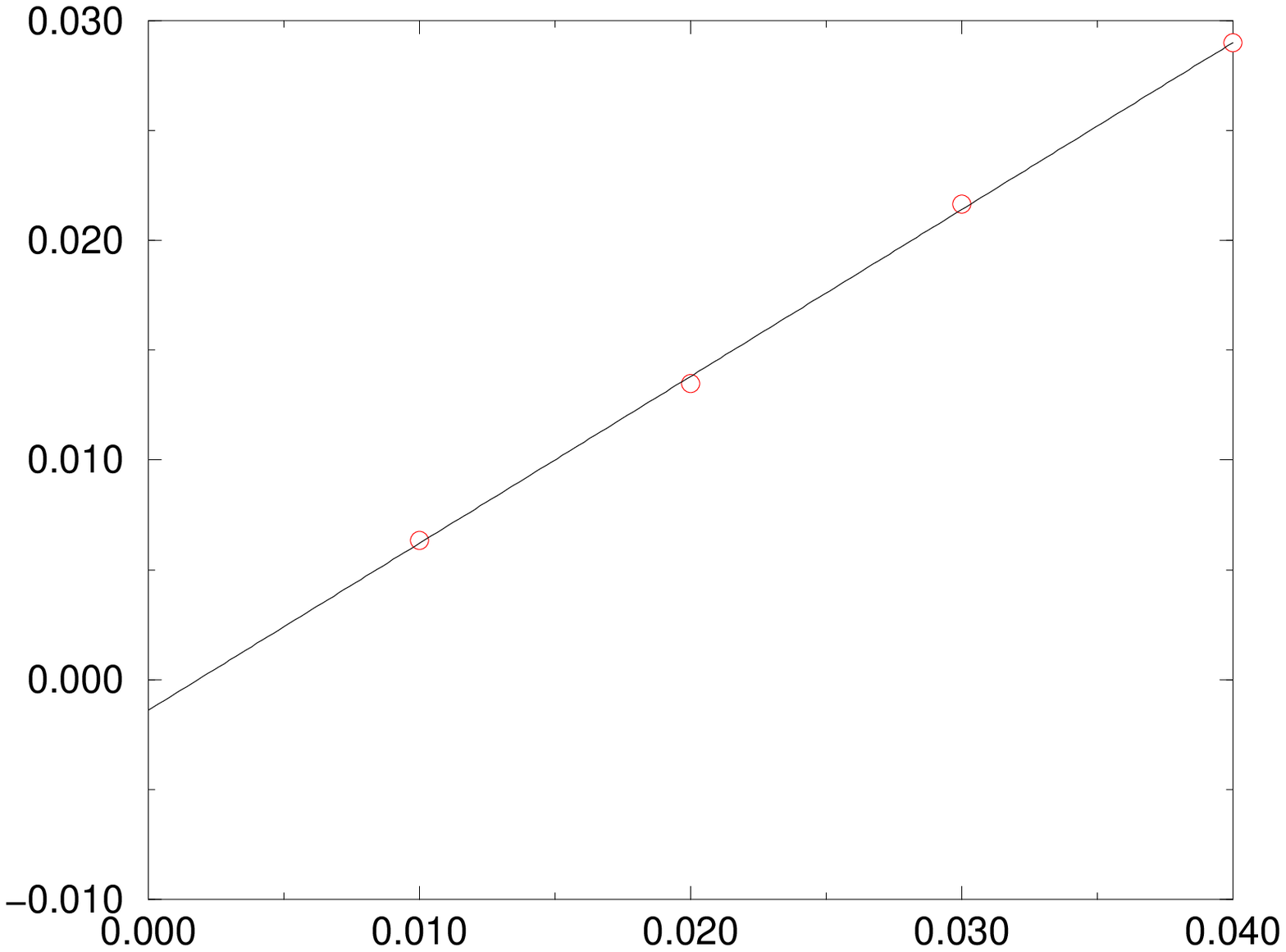}
}
\psfull
\caption{$M_\pi^2\langle \bar\chi \chi\rangle$
vs. $m$, for $\beta=0.65$.
\label{fig:mpicondvm}}
\end{figure}

The pion spectrum provides the most interesting information. In the
broken phase and for vanishing bare mass, the pion is expected to be a
massless Goldstone boson. For non-vanishing bare mass in the broken
phase, the pion mass is related to the chiral condensate by a chiral
Ward identity, plus the assumption of one-pole dominance:
\begin{equation}
        M^2_\pi = \frac{Z_\pi}{\langle \bar\chi \chi \rangle} m.
\label{eq:chiralM}
\end{equation}
Fig.~\ref{fig:mpivm} displays the behaviour of the pion mass squared
vs. $m$. There is a satisfactory agreement with the linear behaviour
expected from Eq.~(\ref{eq:chiralM}). A linear fit to the form:
\begin{equation}
        M^2_\pi = \alpha m + M_0
\end{equation}
yields for the pion mass in the chiral limit $M_0 = 0.008 \pm 0.005$,
which is within two standard deviations from the expected vanishing
value.
Actually, Eq.~(\ref{eq:chiralM}) predicts that the product of the pion
mass squared times the chiral condensate is a linear function of the
bare mass. The product is shown in Fig.~\ref{fig:mpicondvm}. A linear
fit gives a slightly negative intercept for the value of the product
as $m=0$.

Finally in this subsection we discuss the spectroscopy in spin-1 channels.
As described in \cite{DHM}, we performed measurements using both local
and conserved (ie. one-link) operators. In each case the signal 
observed has a strong oscillatory component, being close to or consistent with
zero on even timeslices, suggesting there are light states
in both direct and alternating channels. This motivates the following fitting form:
\begin{equation}
C_v(t)=A(e^{-M_dt}+e^{-M_d(L-t)})+B(-1)^t(e^{-M_at}+e^{-M_a(L-t)}),
\end{equation}
which in general has four parameters. We found, however, that
where fits were possible, in the symmetric phase, the two parameter fit
obtained by setting $A=B$ and $M_d=M_a$ was equally plausible. For values of 
$1/g^2\leq0.8$ the signal to noise ratio rapidly decreased and no sensible fits
could be obtained for $1/g^2\leq0.7$.
The results are given in Table \ref{tab:spec_vect}.

A couple of remarks about the spin-1 sector in 2+1 dimensions are worth making.
First, the spin/flavor assignments are different for the two types of bilinear, 
the local operator projecting onto 
$(\gamma_\mu\gamma_0\otimes\tau_3^*\tau_\mu)$ in direct and 
$(\gamma_5\gamma_\mu\gamma_0\otimes\tau_3^*\tau_\mu)$ in alternating channels
respectively, and the conserved onto $(\gamma_\mu\otimes\One)$ in direct
and $(\gamma_5\gamma_\mu\otimes\tau_3^*)$ in alternating channels, where 
$\mu$ labels one of the transverse directions, the first component of the 
tensor product acts on the four spin indices of the continuum spinor, and
the second on a two-component flavor structure \cite{BB}. It can be seen that 
in each case the direct and alternating channels correspond to states of 
opposite parity, where in 2+1 dimensions in the $q$-basis the parity
transformation $P$ is defined:
\begin{eqnarray}
x=(x_0,x_1,x_2) & \mapsto & x^\prime=(x_0,-x_1,x_2)\nonumber \\
q(x)\mapsto(\gamma_1\gamma_5\otimes\One)q(x^\prime) & ; &
\bar q(x)\mapsto\bar q(x^\prime)(\gamma_5\gamma_1\otimes\One).
\end{eqnarray}
This brings us to the second point: in 2+1 dimensions the angular momentum
operator $J$, which has integer eigenvalues $j$, anticommutes with $P$.
Therefore the two parity eigenstates 
$\vert j,\pm\rangle=\vert j\rangle\pm P\vert
j\rangle$ correspond to distinct non-null eigenstates of the Hamiltonian,
since $\vert j\rangle$ and $P\vert j\rangle$ have distinct $J$ eigenvalues
$j$ and $-j$. Since $P$ commutes with $H$, it follows that if parity is not
spontaneously broken, the model must contain
degenerate states in the massive $j\not=0$ sector of the spectrum. 
Such a parity doubling is observed in the glueball spectrum of
SU($N$) lattice gauge theory in 2+1 dimensions \cite{Mike}.
In the current context we attribute the degeneracy of $M_d$
and $M_a$, as revealed by the quality of the fits of Table \ref{tab:spec_vect},
to parity doubling.
It is not clear how parity doubling is revealed in the $1/N_f$ expansion, 
which at leading order predicts a massive bound state in the 
$\gamma_\mu$ channel but not in the 
$\gamma_5\gamma_\mu$ channel \cite{DHM}.

\section{$N_f=6$}

\begin{figure}[htb]
\psdraft
\centerline{
\setlength\epsfxsize{300pt}
\epsfbox{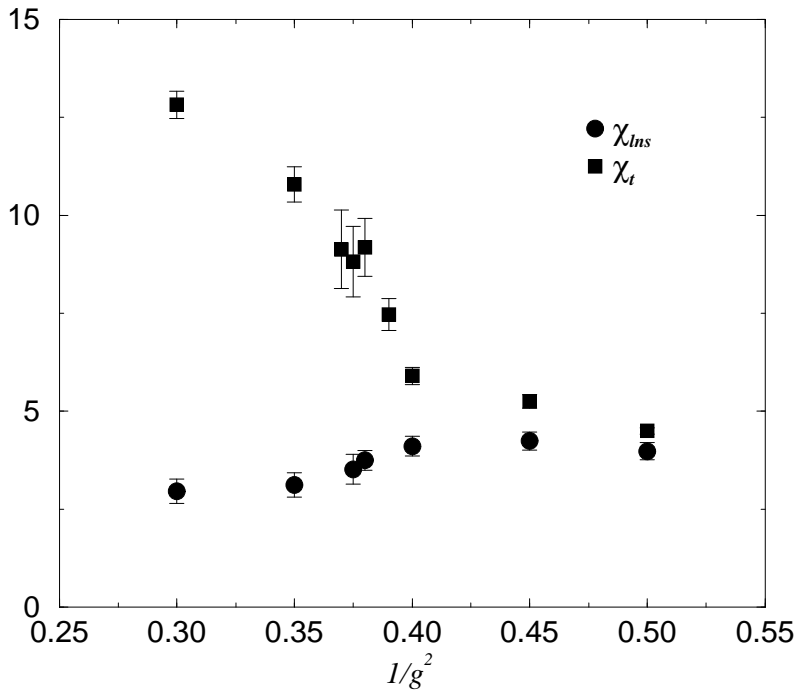
}}
\psfull
\caption{ Plot of $\chi_t$ and $\chi_{nls}$ versus $1/g^2$ for $N_f=6$
on a $16^3$ lattice with $m=0.01$.
\label{fig:n=3susc}}
\end{figure}

\begin{figure}[htb]
\psdraft
\centerline{
\setlength\epsfxsize{300pt}
\epsfbox{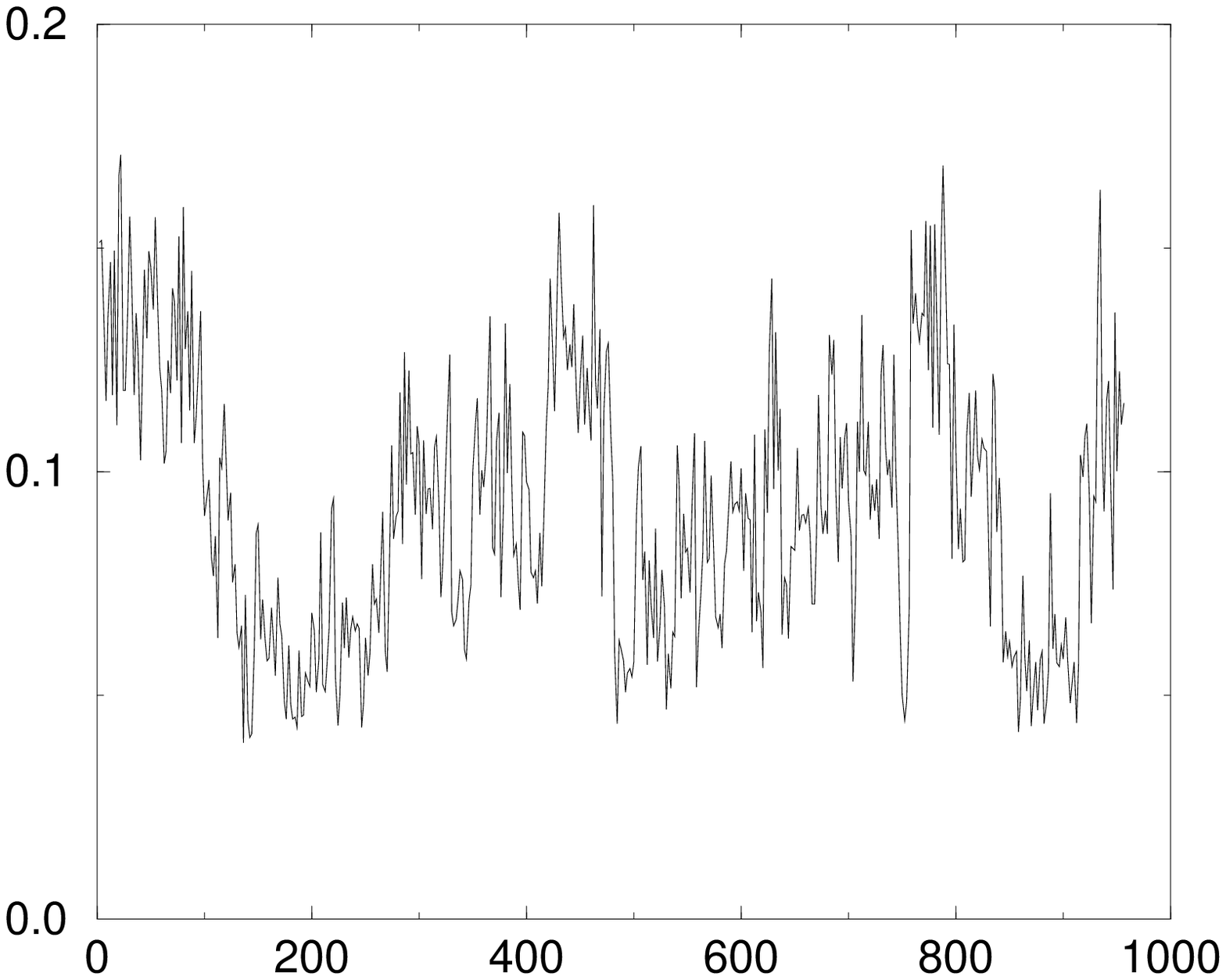
}}
\psfull
\caption{Time history for chiral condensate measurements
for $N_f=6$ on a $16^3$ lattice with $m=0.01$ and $1/g^2=0.38$.
\label{fig:hist38}}
\end{figure}

\begin{figure}[htb]
\psdraft
\centerline{
\setlength\epsfxsize{300pt}
\epsfbox{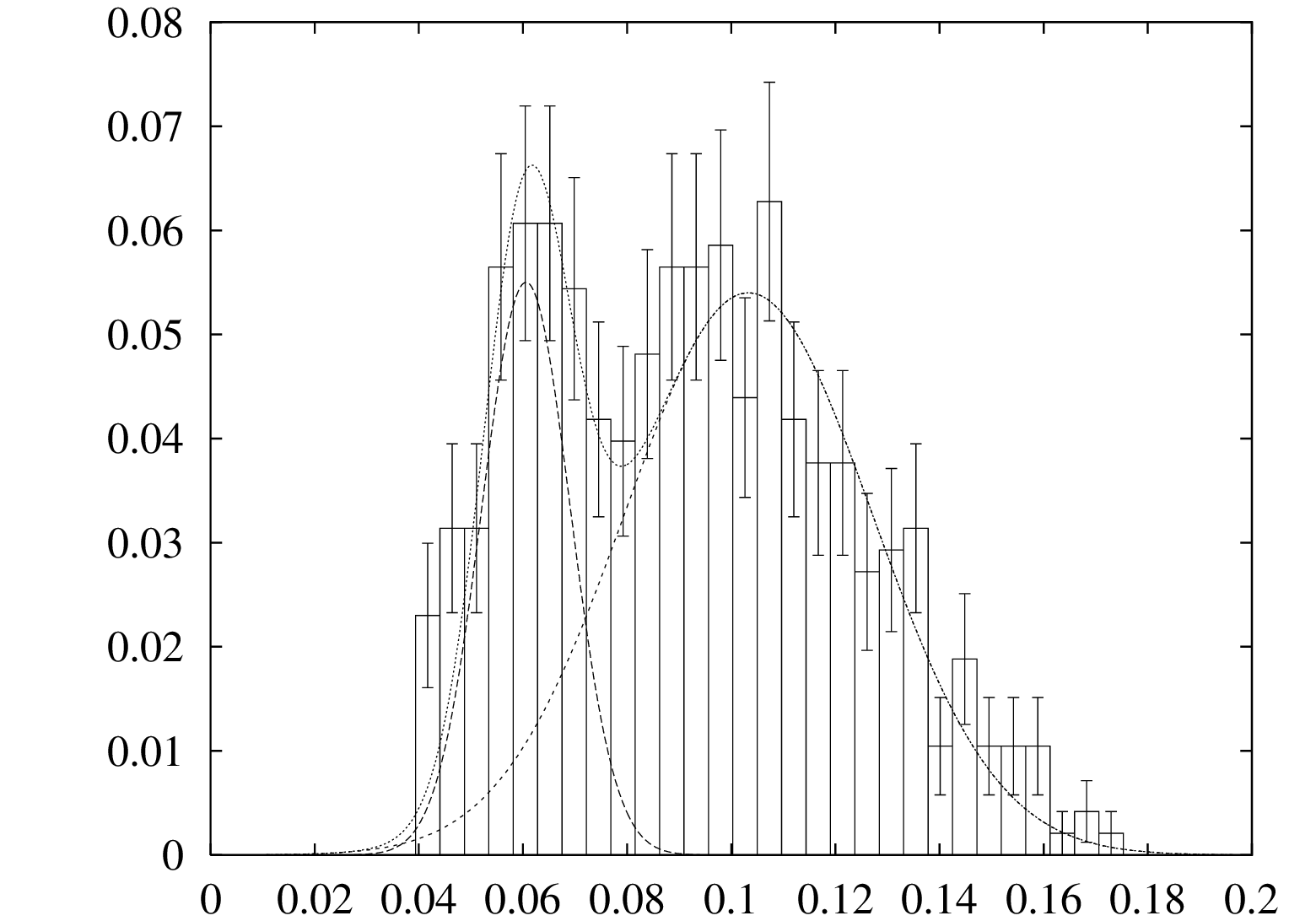
}}
\psfull
\caption{Histogram of 478 chiral condensate measurements for
$N_f=6$ on a $16^3$ lattice with $m=0.01$, $1/g^2=0.38$,
together with a fit to a double gaussian ($\chi^2/\mbox{dof}=32/26$). 
\label{fig:twinpeak}}
\end{figure}

\begin{figure}[htb]
\psdraft
\centerline{
\setlength\epsfxsize{300pt}
\epsfbox{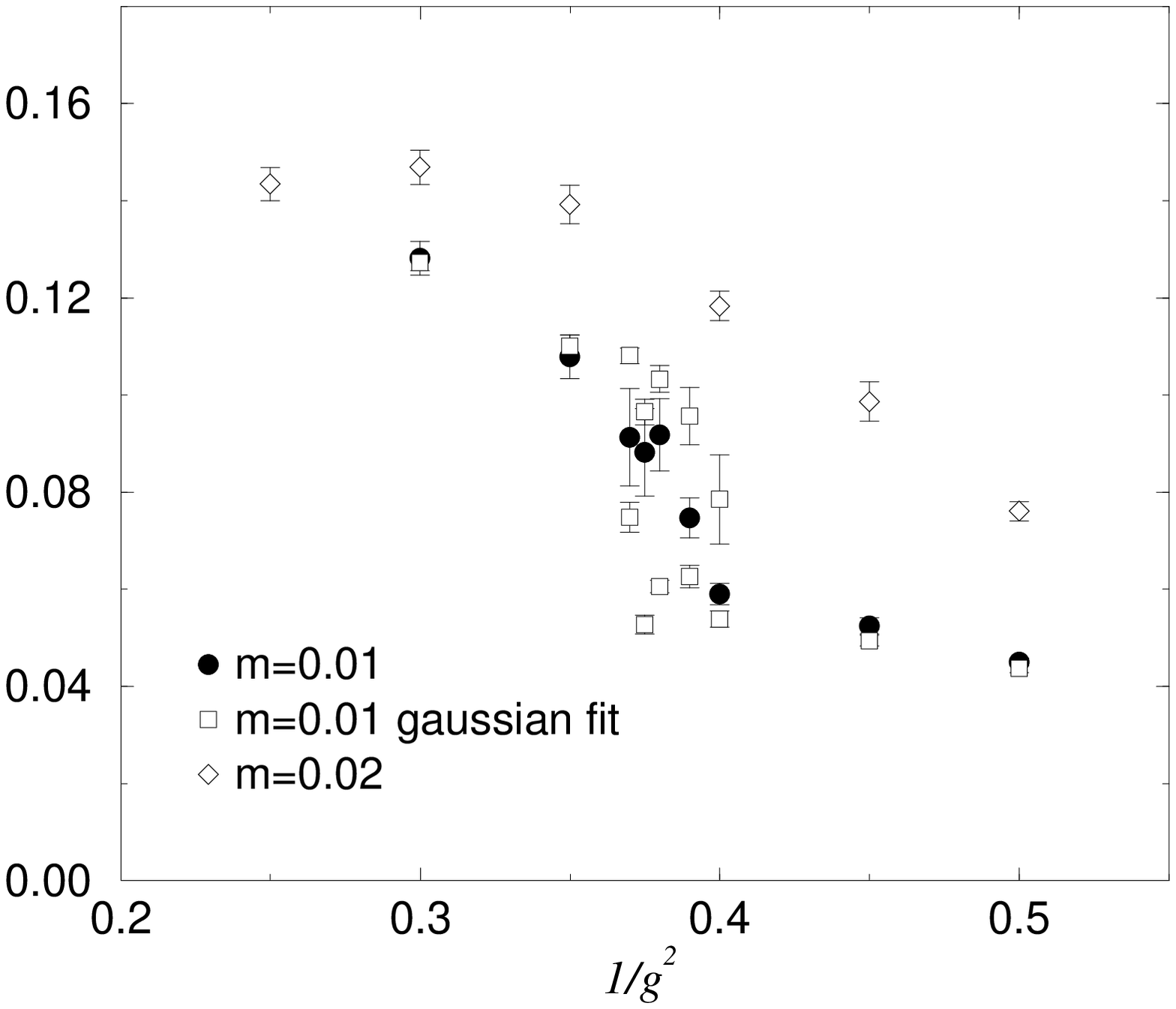
}}
\psfull
\caption{Chiral condensate versus $1/g^2$ for
$N_f=6$ on a $16^3$ lattice, showing
evidence for coexisting phases for $m=0.01$ based on
double gaussian fits. 
\label{fig:n=3cond}}
\end{figure}

In this section we turn our attention to the case $N_f=6$. Although
we have spent a comparable effort in accumulating data for this case, 
the runs are
more compute intensive, firstly for the obvious reason that an extra 
lattice flavor implies an extra matrix inversion, and secondly because 
the physically interesting regime occurs at 
still stronger coupling -- we explored the region $1/g^2\in[0.3,0.5]$,
again with lattice size $16^3$ and $m=0.01$.
Therefore the discussion will be more qualitative. Our primary observation is
that there is still evidence for chiral symmetry breaking at strong
coupling, but that the phase transition is of a different character.
The evidence for a phase transition is shown in Fig. \ref{fig:n=3susc},
where a comparison between tranverse and longitudinal 
non-singlet susceptibilities shows a 
clear separation between a chirally symmetric region
($\chi_t\simeq\chi_{lns}$) for $1/g^2>0.4$ and a chirally broken region
($\chi_t\gg\chi_{nls}$) for $1/g^2<0.35$. Consider, however,
the time history of condensate measurements obtained 
at $1/g^2=0.38$ from 
nearly 1000 trajectories of mean length 0.9, shown in Fig. \ref{fig:hist38}.
It is plausible that the figure shows evidence for metastability, the system
tunnelling between equilibrium states with
$\langle\bar\chi\chi\rangle\simeq0.05$ and 
$\langle\bar\chi\chi\rangle\simeq0.1$. A comparison with a similar plot 
for the susceptibility ratio $\chi_t/\chi_{lns}$ reveals a correlation: 
when the condensate is small the ratio is close to one, suggesting 
that chiral symmetry is realised, whereas when the
condensate is larger the ratio is much smaller, with occasional excursions to
negative values, similar to the broken phase behaviour of Fig.
\ref{fig:history}. This encourages us to interpret the time history as 
a sequence of tunnellings between chirally symmetric and broken vacua.
A histogram of the condensate measurements is shown in Fig. \ref{fig:twinpeak},
together with a double gaussian fit of the form 
\begin{equation}
y=A_1\exp\left(-{(x-\bar x_1)^2\over{2\sigma_1^2}}\right)
+A_2\exp\left(-{(x-\bar x_2)^2\over{2\sigma_2^2}}\right).
\label{eq:dg}
\end{equation}
The data are consistent with the 
twin peak structure characteristic of coexisting
phases.

In Fig. \ref{fig:n=3cond} we plot $\langle\bar\chi\chi\rangle$ versus
$1/g^2$, for $m=0.01$ and $m=0.02$ (this data taken from \cite{DHM}).
The $m=0.01$ data is plotted in two ways, firstly as a raw average over the
whole dataset, and secondly by assuming coexistent states, with central 
values and standard errors taken from fits to (\ref{eq:dg}). We found that
the two procedures yield consistent results for $1/g^2\leq0.35$ and
$1/g^2\geq0.45$; however for intermediate couplings there is a marked 
twin peak structure and evidence for coexisting states.

\begin{figure}[htb]
\psdraft
\centerline{
\setlength\epsfxsize{300pt}
\epsfbox{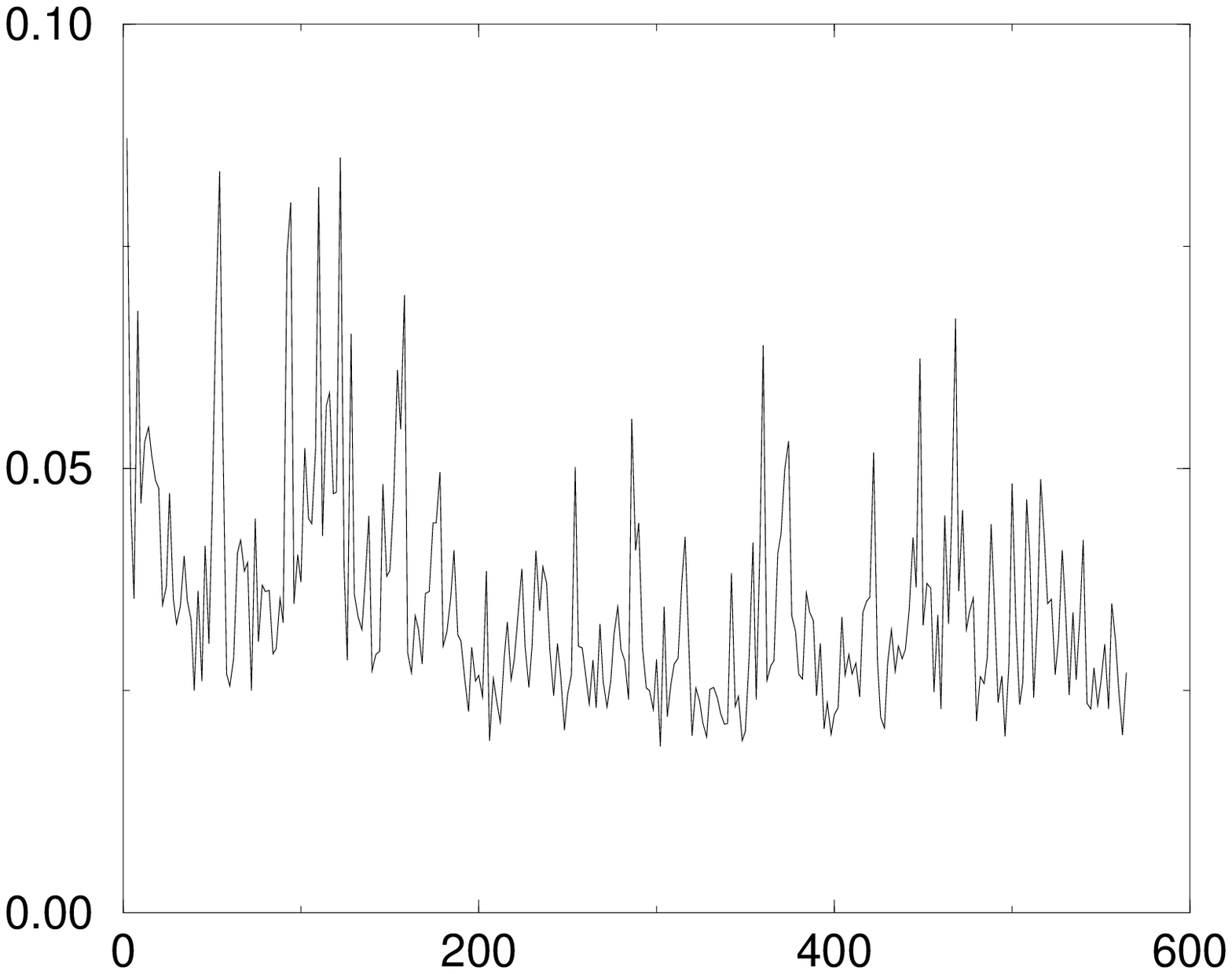
}}
\psfull
\caption{Time history for chiral condensate measurements
for $N_f=6$ on a $16^3$ lattice with $m=0.005$ and $1/g^2=0.37$.
\label{fig:hist37}}
\end{figure}

In ref. \cite{DHM} we presented data for $m=0.05,\dots,0.02$ but observed
no evidence for a critical point described by a Fisher plot, consistent 
with a second order phase transition. The new data from a larger lattice
closer to the chiral limit suggest that there is indeed a phase transition,
but that it is first order, as signalled by the evidence for 
coexisting phases in the transition region. 
To confirm this, of course, data from a variety 
of lattice volumes and bare fermion masses would be needed, requiring
resources beyond the scope of the current study. We did attempt to study 
the transition closer to the chiral limit with some runs at $m=0.005$; a typical
time history is shown in Fig. \ref{fig:hist37}. The results were ambiguous, with
no clear evidence for a chirally broken phase at any coupling studied down to 
$1/g^2=0.36$; this is a sure 
sign that any conclusions made are as yet preliminary.

\section{Discussion}

The main results we have found are that 
the $N_f=4$ Thirring model in three dimensions,
like its $N_f=2$ counterpart, appears to have a continuous chiral symmetry
breaking transition at strong coupling, whereas for the $N_f=6$ case the
data suggests that the transition is first order. Moreover, the $N_f=4$
model has critical exponents distinct from those of $N_f=2$, and we
expect this conclusion to be robust even if the actual numerical values of the
exponents drift somewhat under more refined analyses in the future.
The conclusion we draw is that both $N_f=2$ and $N_f=4$ models have 
interacting continuum limits, which are qualitatively, but not quantitatively
similar. This is consistent with the critical flavor number $N_{fc}$, predicted
in the strong coupling limit of the Schwinger-Dyson approach to the model
to be approximately 4.32
in ref. \cite{Itoh}, to be at least greater than 4. For $N_f=6$, on the other
hand, there can be no continuum limit. It is interesting to compare our
results with those found in simulations of $\mbox{QED}_4$; power-law fits to the
equation of state \cite{Rpi}\cite{QED4} 
yield critical exponents which are not distinct
for $N_f=2, 4$, but studies at larger $N_f$ do suggest that the chiral 
transition eventually becomes first order in this limit \cite{QED4N}.

It is difficult at first sight 
to reconcile the $N_f=6$ result with the Schwinger-Dyson
picture, which for $N_f>N_{fc}$ would simply predict the absence of a stable
chiral symmetry breaking solution for any finite coupling. However, as explained
in \cite{DHM}, the lattice regularisation of the Thirring model itself contains
systematic uncertainties, essentially because the interaction current 
in the contact term is not actually conserved. In the context of the
$1/N_f$ expansion, this results in an uncancelled linear divergence in the
calculation of the vacuum polarisation tensor, which controls the propagation
of the $f\bar f$ state in the vector channel, and which must be  absorbed by an 
additive renormalisation of the lattice inverse coupling constant:
\begin{equation}
{1\over g_R^2}={1\over g^2}-J(m),
\end{equation}
where $J(m)$ is calculable to leading order in $1/N_f$. Therefore we might
expect the strong coupling limit to be attained for $1/g^2=J(m)$; indeed for 
$1/g^2<J(m)$ the $1/N_f$ expansion of the lattice model is not unitary.
What the results in this paper show is that the true picture at small $1/g^2$
is not described by the $1/N_f$ expansion, probably because it assumes the
wrong vacuum state, ie. one where chiral symmetry is realised.
Instead, at strong couplings, or equivalently for sufficiently many fermion
flavors, the chiral transition becomes first order, invalidating approaches 
such as either the $1/N_f$ expansion or the Schwinger-Dyson equation, which
both rely on the applicability of continuum field theory. We can, however, 
retain the notion of a critical flavor number $N_{fc}$, such that chiral
symmetry breaking solutions
such as those of \cite{Itoh} are applicable, and RG fixed points exist 
for $N_f<N_{fc}$. In this case our main result can be stated:
\begin{equation}
4<N_{fc}<6.
\end{equation}
Work is in progress, using a generalisation of the numerical algorithm
to allow for non-integer $N_f$, 
to explore further the phase diagram of the model in 
the $(1/g^2, N_f)$ plane, and hopefully to pin down the value of $N_{fc}$
more precisely \cite{Biagio}.

Finally, what of the theory with $N_f\simeq N_{fc}$? According to \cite{Itoh}, 
at this point we expect the induced physical scale 
(inverse correlation length) $\mu$
to scale in an essentially singular way:
\begin{equation}
{\mu\over\Lambda}\propto\exp\left(-{{2\pi}\over\sqrt{{N_{fc}\over N_f}-1}}
\right),
\end{equation}
a form of scaling characteristic of a conformal phase transition
\cite{MirYam}. Indeed, following the discussion of the introduction, it
is precisely for $N_f=N_{fc}$ that we might expect the Thirring model
and $\mbox{QED}_3$ to coincide \cite{Gusy}\cite{DHM}.
On the assumption that this limit might be approached along a smooth
trajectory corresponding to a line of phase transitions 
in the $(1/g^2,N_f)$ plane, and that critical exponents vary continuously
along this line, then we have an apparent contradiction; our numerical results
suggest that the exponent $\delta$ {\sl increases\/} 
as $N_f\nearrow N_{fc}$, whereas
for a chiral transition described by essential singularity, such as an
asymptotically-free
theory, or the quenched gauged NJL model \cite{DHKK}, $\delta$ is expected to
take the value 1, and hence {\sl decrease\/} as this limit is approached.
Therefore we conclude that our simulation results do not appear to support the
existence of a conformal phase transition. 

\section{Acknowledgements}

SJH was supported in part by a PPARC Advanced Fellowship, and in part
by the TMR-network ``Finite temperature phase transitions in particle
physics'' EU-contract ERBFMRX-CT97-0122. LDD is supported by PPARC
under grants GR/L56329 and GR/L29927. Some of the computing work was
performed using the resources of the UKQCD collaboration under PPARC
grants GR/K41663, GR/K455745 and GR/L29927. We have enjoyed discussing
aspects of this work with Biagio Lucini, Ji\v r\'\i ~Jers\'ak and
Wolfgang Franzki.

\clearpage

\end{document}